\documentclass[10pt,journal,compsoc,twoside]{IEEEtran}

\ifCLASSOPTIONcompsoc
  \usepackage[nocompress]{cite}
\else
  \usepackage{cite}
\fi

\ifCLASSINFOpdf
\else
\fi

\usepackage{amssymb}
\usepackage{amsmath}
\usepackage{lineno}
\usepackage[ruled]{algorithm2e}
\usepackage[hyphens]{url}
\usepackage{color}
\usepackage{mathtools}
\usepackage{lineno}
\usepackage{array}
\usepackage{pdflscape}
\usepackage{afterpage}
\usepackage{multirow}
\usepackage{booktabs}
\usepackage{url}
\usepackage{balance}
\usepackage[pdfencoding=auto]{hyperref}

\expandafter\def\expandafter\UrlBreaks\expandafter{\UrlBreaks%  save the current one
  \do\a\do\b\do\c\do\d\do\e\do\f\do\g\do\h\do\i\do\j%
  \do\k\do\l\do\m\do\n\do\o\do\p\do\q\do\r\do\s\do\t%
  \do\u\do\v\do\w\do\x\do\y\do\z\do\A\do\B\do\C\do\D%
  \do\E\do\F\do\G\do\H\do\I\do\J\do\K\do\L\do\M\do\N%
  \do\O\do\P\do\Q\do\R\do\S\do\T\do\U\do\V\do\W\do\X%
  \do\Y\do\Z}

\usepackage{rotating}
\usepackage{amsmath}
\usepackage[nodayofweek]{datetime}
\newdateformat{mytoday}{\monthname[\THEMONTH] \twodigit{\THEDAY}, \THEYEAR}
\usepackage{mathtools}
\usepackage{array}
\usepackage{tabulary}
\newcolumntype{K}[1]{>{\centering\arraybackslash}p{#1}}

\makeatletter
\long\def\@IEEEtitleabstractindextextbox#1{\parbox{0.922\textwidth}{#1}}
\makeatother

\IEEEoverridecommandlockouts

\begin{document}

%\noindent Cite this preprint version of the manuscript as:
%
%\ \\
%\noindent \textcolor{blue}{F. Afsana, M.A. Rahman, M.R. Ahmed, M. Mahmud, M.S. Kaiser, ``An Energy Conserving Routing Scheme for Wireless Body Sensor Nanonetwork Communication,'' \textit{IEEE Access}, 2018, doi: 10.1109/ACCESS.2018.2789437.
%\ \\
%\noindent \textcopyright\ IEEE holds the copyright of this work.
%}

%\IEEEpubid{\begin{minipage}{\textwidth}\ \\[12pt]
% 978-1-5386-0869-2/17/\$31.00 \copyright 2017 IEEE \\
% 4th International Conference on Advances in Electrical Engineering 28-30 September, 2017, Dhaka, Bangladesh
%\end{minipage}}

\IEEEpubid{\begin{minipage}{\textwidth}\ \\[12pt]
\textcolor{blue}{Cite as: F. Afsana, M.A. Rahman, M.R. Ahmed, M. Mahmud, M.S. Kaiser, ``An Energy Conserving Routing Scheme for Wireless Body Sensor Nanonetwork Communication,'' \textit{IEEE Access}, 2018, doi: 10.1109/ACCESS.2018.2789437. \copyright\ 2018 IEEE.}
\end{minipage}}

\title{An Energy Conserving Routing Scheme for Wireless Body Sensor Nanonetwork Communication}
%\IEEEaftertitletext{\textcolor{blue}{Cite as: F. Afsana, M.A. Rahman, M.R. Ahmed, M. Mahmud, M.S. Kaiser, ``An Energy Conserving Routing Scheme for Wireless Body Sensor Nanonetwork Communication,'' \textit{IEEE Access}, 2018, doi: 10.1109/ACCESS.2018.2789437. \textcopyright\ IEEE.}}

\author{Fariha Afsana,~\IEEEmembership{Student Member,~IEEE,} Md. Asif-Ur-Rahman, Muhammad R. Ahmed,~\IEEEmembership{Member,~IEEE,}  Mufti~Mahmud$^1$,~\IEEEmembership{Senior Member,~IEEE,}
        and M~Shamim~Kaiser$^1$,~\IEEEmembership{Senior Member,~IEEE,}
\thanks{$^1$ corresponding authors. Emails: muftimahmud@gmail.com (M. Mahmud), mskaiser@juniv.edu (M. S. Kaiser)}
\thanks{F. Afsana is with the Institute of Information Technology, Jahangirnagar University, Savar, 1342 - Dhaka, Bangladesh}% <-this % stops a space

\thanks{M. Arif-Ur-Rhaman is with the Department of Computer Science, American International University-Bangladesh, 1213 - Dhaka, Bangladesh}% <-this % stops a space

\thanks{M. R. Ahmed is with the Radio, Radar and Communications, Military Technological College, Muscat, Oman}% <-this % stops a space

\thanks{M. Mahmud is with the NeuroChip Lab, Department of Biomedical Sciences, University of Padova, Via F. Marzolo 3, 35131 - Padova, Italy}% <-this % stops a space
\thanks{M. S. Kaiser is with the Institute of Information Technology, Jahangirnagar University, Savar, 1342 - Dhaka, Bangladesh}% <-this % stops a space

\thanks{Version: \mytoday \today; last saved: \currenttime~(GMT)}
}

\markboth{Afsana \MakeLowercase{\textit{et al.}}: An Energy Conserving Routing Scheme for WBSN Communication}%
{Afsana \MakeLowercase{\textit{et al.}}: An Energy Conserving Routing Scheme for WBSN Communication}

\IEEEtitleabstractindextext{%
\begin{abstract}
Current developments in nanotechnology make electromagnetic communication possible at the nanoscale for applications involving Body Sensor Networks (BSNs). This specialized branch of Wireless Sensor Networks, drawing attention from diverse fields such as engineering, medicine, biology, physics and computer science, has emerged as an important research area contributing to medical treatment, social welfare, and sports. The concept is based on the interaction of integrated nanoscale machines by means of wireless communications. One key hurdle for advancing nanocommunications is the lack of an apposite networking protocol to address the upcoming needs of the nanonetworks. Recently, some key challenges have been identified, such as nanonodes with extreme energy constraints, limited computational capabilities, Terahertz frequency bands with limited transmission range, etc., in designing protocols for wireless nanosensor networks. This work proposes an improved performance scheme of nanocommunication over Terahertz bands for wireless BSNs making it suitable for smart e-health applications. The scheme contains -- a new energy-efficient forwarding routine for electromagnetic communication in wireless nanonetworks consisting of hybrid clusters with centralized scheduling; a model designed for channel behavior taking into account the aggregated impact of molecular absorption, spreading loss, and shadowing; and an energy model for energy harvesting and consumption. The outage probability is derived for both single and multilinks and extended to determine the outage capacity. The outage probability for a multilink is derived using a cooperative fusion technique at a predefined fusion node. Simulated using a Nano-Sim simulator, performance of the proposed model has been evaluated for energy efficiency, outage capacity, and outage probability. The results demonstrate the efficiency of the proposed scheme through maximized energy utilization in both single and multihop communication; multisensor fusion at the fusion node enhances the link quality of the transmission.

%\ \\
%\textcolor{blue}{Cite as: F. Afsana, M.A. Rahman, M.R. Ahmed, M. Mahmud, M.S. Kaiser, ``An Energy Conserving Routing Scheme for Wireless Body Sensor Nanonetwork Communication,'' \textit{IEEE Access}, 2018, doi: 10.1109/ACCESS.2018.2789437. \textcopyright\ IEEE.}
\end{abstract}

}

\maketitle

\IEEEdisplaynontitleabstractindextext
%\IEEEpeerreviewmaketitle

\ifCLASSOPTIONcompsoc
\IEEEraisesectionheading{\section{Introduction}\label{sec:introduction}}
\else
\section{Introduction}
\label{sec:introduction}
\fi

\IEEEPARstart{R}{ecent} advances in nanotechnology expanded its potential applications in diverse fields. In the Biomedical field Nanotechnology has been successfully applied in healthcare, health monitoring, support immune systems, creating novel drug delivery systems, detecting the presence of infectious agents, and identifying specific types of cancer \cite{yamada2003,sahoo2007,wang2012}. It has been used in waste-water management, controlling biodiversity, and assisting biodegradation to improve the environment \cite{gao2014}. Nanomaterials play important roles in industrial manufacturing processes, quality control procedures, and self-cleaning anti-microbial textiles \cite{5akyildiz2008,6rikhtegar2013}. `Nanomachines' have also been deployed in nuclear, biological and chemical defense systems for surveillance for military and defense purposes \cite{akyildiz2011}.
A nanomachine is composed of a power supply, memory, antenna and CPU module, and it behaves as an autonomous node capable of performing tasks such as computing, storing, sensing and/or actuating at the nanolevel. 
The nanomachines are fabricated using three approaches - top-down, bottom-up, and bio-hybrid approach \cite{1gine2009}. 
Existing electronic devices are scaled down to the nanolevel using the top-down approach, but in the case of bottom-up, molecular components are used to create the nanomachines. The bio-hybrid approach uses biological elements as patterns to be replicated with synthesized components \cite{2haselman2010}.

The nanomachines with limited capabilities perform tasks by forming an interconnected nanonetwork. The traditional communication protocols are not applicable to enable the nodes to communicate among themselves at the nanoscale, these conventional systems need to undergo extensive revisions. Based on the preferred transmission medium, different communication paradigms have been proposed over time, which include acoustic, nanomechanical, molecular, and electromagnetic (EM) \cite{7akyildiz2014}.

For practical reasons, only specific paradigms can be applied to specific scenarios. For example, due to a very high absorption coefficient of acoustic waves in bones and muscle fiber, using acoustic paradigm based communication is impractical \cite{5akyildiz2008}. Nanomechanical communication requires physical contact between transmitting and receiving nanomachines, which is not always possible. The molecular and EM communication paradigms have been very promising  \cite{7akyildiz2014} with molecules as the encoder, transmitter and receiver of information in molecular communication 
\cite{malak2012} and EM waves as the communication means among nanomachines in EM communication \cite{14akyildiz2010}.

Networks operating in the Terahertz (THz) band (i.e., $0.06 - 10$ THz) offer promising solutions for many applications, including health monitoring, biochemical weapon monitoring, and plant monitoring, which otherwise are not possible \cite{7akyildiz2014}. Recent advances in THz communication allow development of a new generation of nano-EM communication schemes. Because many molecules resonate within the THz band, THz radiation may exploit this to penetrate opaque materials with high selectivity \cite{8moldovan2014}, making it suitable for biomedical and healthcare applications \cite{piro2016}.

The need for energy-efficient, low cost, simple and scalable healthcare solutions is at its peak and can be served with miniaturized devices (MD) \cite{patel2010}. 
However, existing MD systems have difficulty performing multiple tasks simultaneously, i.e., functions offered by consumer electronic devices, and functions related to healthcare (e.g., long-term health monitoring). The wireless body sensor network (WBSN) can achieve this objective by providing inexpensive and continuous health monitoring by implanting tiny biosensors inside the human body, which do not impair normal activities  \cite{abbasi_2016_nanocom_review}. These devices then communicate among themselves, as required, and transmit the required information to a receiving end. To make this communication smooth, different strategies have been investigated \cite{cavallari2014}. In particular, due to the short communication range of the nanodevices, a multihop EM communication paradigm has been extensively explored with appropriate EM power \cite{abbasi_2016_nanocom_review}. Some late methods addressing communication at the nanolevel include Graphene-enabled EM communication, molecular diffusion, Foster Resonance Energy Transfer \cite{10kuscu2011}, etc. However, direct applicability at the nanoscale is hampered by device size, complexity, and energy consumption \cite{misra2014}. Additionally, the large bandwidth of the THz band (i.e., Terabits per second, Tbps) is susceptible to shadowing and noise \cite{12dressler2015}.

This work proposes a lightweight communication scheme suitable for WBSN operating at the THz band with four main contributions:\\
i) First, an energy-efficient forwarding scheme based on Nano Cluster Composition Algorithm (NCCA) is proposed for THz band communication in WBSN.\\
ii) Second, a channel behavior model is investigated on the basis of the aggregated impact of molecular absorption, spreading losses, and shadowing.\\
iii) Third, the joint impact of energy harvesting and energy consumption processes is considered for developing a nanosensor energy model, which includes the stochastic nature of energy arrival.\\
iv) Finally, closed-form outage probability expressions are deduced for both single and multilinks. To enhance information detection precision and energy-efficiency, in the case of multilink, a cooperative fusion approach is used where data from disparate sources are fused at a predefined fusion node. Given the outage probability, the approach has been extended to determine outage capacity of the considered channel.

The remainder of the paper is organized as follows. Related work on existing schemes is presented in section \ref{sec-sota}. The system model is presented in \ref{sec-systmodel}, which includes the channel behavior model and nanosensor energy model. Section \ref{sec-perfass} is devoted to performance assessment of the proposed scheme from simulation results. In addition, final concluding remarks are provided in section \ref{sec-conc}.

\section{EM Communication at the Nanoscale}
\label{sec-sota}
EM communication has attracted increasing attention due to its suitability in nanonetworks. EM communication at the nanoscale has been reviewed from different perspectives highlighting various opportunities and many associated challenges. In an in-depth review, Akyildiz and Jornet discussed EM communication in nanonetworks, identified several communication challenges at the nanoscale (e.g., THz channel modeling), and proposed a roadmap \cite{14akyildiz2010}. Yu et al. provided the available forwarding schemes for EM-based wireless nanosensor networks (WNSN) in the THz band \cite{Yu_2015_em_wnsn}.

\subsection{EM Communication and the Internet-of-Nano-Things}
\label{subsec-iont}
The `Internet of Things' has also been extended to the `Internet of Nano Things' (IoNT) by interconnecting nanonetworks to the Internet. A reference architecture of the IoNT was introduced in  \cite{11akyildiz2010} by Akyildiz and Jornet where the THz band was considered for EM communication at the nanoscale. Few major challenges concerning channel modeling, information modulation, and protocols for nanonetworks were highlighted. In another study, Dressler and Fischer investigated connecting a body area network with in-body nanodevices \cite{12dressler2015}, identified the challenges of such integration, proposed associated network architecture, and laid the foundation for simulation-based performance evaluation of such concepts. A WNSN, formed by hexagonal cell-based nanosensors deployed randomly at different body organs, was proposed in \cite{13lee2015} for intra-body disease detection as an application of the IoNT.

\subsection{Network Communication Protocols}
\label{subsec-netcommprot}
A routing framework considering the peculiarities of the WNSNs was proposed by Pierobon et al., both in terms of nanoscale energy harvesting process and Terahertz band wireless communication \cite{17pierobon2014}. A hierarchical cluster-based architecture was considered to reduce network operation complexity. The proposed routing framework was based on a previously proposed medium access control (MAC) protocol where joint throughput and lifetime scheduling were performed by a nanocontroller  on top of a time division multiple access (TDMA) framework within each cluster. Selecting efficient multihop communication saved energy. The total routing strategy was conceived over a cluster and communication of nanosensors within it. How the clusters were constructed and how inter-cluster communication occurred was not clear.

Based on a WNSN using EM communication in the THz band, Piro et al. proposed a new routing algorithm along with an efficient MAC protocol and improved Nano-Sim (see \cite{pironanosim2013} for nano-sim), an NS-3 module to model WNSNs \cite{20piro2013}. 

A hierarchical network architecture integrating Body Area Nano-NETwork (BANNET) and an external macroscale healthcare monitoring system was proposed \cite{18piro2015}. In this work, two different energy-harvesting aware protocol stacks using optimal routing protocol and a greedy routing approach were conceived for handling the communication of nanosensors moving uniformly in an environment mimicking human blood vessels. An energy-aware MAC protocol was used in both strategies to identify the available nanonodes through a handshake mechanism. These two schemes performed well compared with the simple flooding scheme. However, high computational capacity was required for the optimal scheme.

A channel aware forwarding scheme for EM-based WNSN (EM-WNSN) in the THz band was proposed by Yu et al. as a solution to the THz frequency selective feature of networking \cite{Yu_2015_em_wnsn}. This work evaluated classical multihop forwarding and single end-to-end transmission schemes for EM-WNSNs. Jornet et al. proposed a Physical Layer Aware MAC protocol for EM nanonetworks in the THz band (PHLAME) where the transmitting and receiving nanodevices were allowed to jointly select the communication parameters in an adaptive fashion  \cite{21jornet2012}. The protocol minimizes interference in the nanonetwork and maximizes the probability of successfully decoding the received information. Its performance was analyzed in terms of energy consumption per unit bit of information, normalized available throughput and average packet delay. However, due to the usage of a handshake during the transmission process and large computations deployed on nanosensors to dynamically find communication parameters, its wide usage in WNSNs has been hampered. The energy and spectrum-aware MAC protocol was an approach to achieve perpetual WNSNs \cite{22wang2013}. The objective was to achieve fair throughput and lifetime optimal channel access by jointly optimizing energy harvesting and consumption processes in nanosensors. A novel symbol compression scheduling algorithm was proposed based on critical packet transmission ratio. Packet-level timeline scheduling was also proposed to achieve balanced single-user throughput with an infinite network lifetime. The whole scenario assumed that all the nanosensors could communicate directly with the nanocontroller in one hop, which is not feasible.

Mohrehkesh and Weigle proposed a method to maximally utilize the harvested energy for nanonodes in perpetual WNSNs that communicate in the THz band \cite{27mohrehkesh2014}. They designed an energy model as a Markov decision process considering that energy arrivals follow a stochastic process. Additionally, a lightweight heuristic method was proposed for resource-limited nanonodes that could perform close to optimal and would not require a heavy processing load on a nanonode. In addition to the development of sophisticated energy-aware communication protocols, successful WNSN systems should have energy sources to provide sufficient energy for the nanodevices to communicate. Jornet and Akyildiz proposed an energy model for self-powered nanosensors to jointly analyze the energy harvesting and energy consumption processes \cite{26jornet2012}. The energy harvesting process was realized by means of a piezoelectric nanogenerator, and the energy consumption process was reviewed for EM communication in the THz band. A mathematical framework was also developed to obtain the probability distribution of nanosensor energy and to investigate the end-to-end packet delay and achievable throughput of WNSNs. 

Additionally, there are few relevant works proposed for WSNs but applicable to WNSNs with minor modification and extension. Afsar and Tayarani \cite{28afsar2014} surveyed different clustering approaches in WSNs from the viewpoint of the objectives and characteristics of the approaches and classified the available clustering algorithms based on different metrics (e.g., mobility, cluster count, cluster size, and complexity of algorithm). Younis and Fahmy proposed a new energy efficient clustering of sensor nodes in ad hoc sensor networks \cite{29younis2004}, called Hybrid Energy-Efficient Distributed clustering (HEED), which performs uniform cluster head distribution across the network. Afsana et al. evaluated the performance of multihop links and approximated outage probability based on end-to-end signal-to-interference plus noise ratio (SINR) for wireless ad hoc networks \cite{30afsana2014}. To better characterize the performance of channels, Behruzifar and Falahati \cite{31behruzifar2012} proposed another approach, which maximizes the outage capacity over cut-off fading states for wireless channels assuming that both transmitter and receiver have full channel state information (CSI). Yigit et al. proposed channel-aware routing for smart grid applications \cite{Yigit2016}, and Zenia et al. reviewed energy efficient routing protocols for underwater WSN \cite{Zenia_2016_uwsn}.

%%%%%%%%%%%%%%%%%%%%%%%%%%%%%%%%%%%%%%%%%%%%%%%%%%%%
% figure 1 is placed here for its placement at the right location in the article

\begin{figure*}[htb!]
 \centering
      \includegraphics[scale=1]{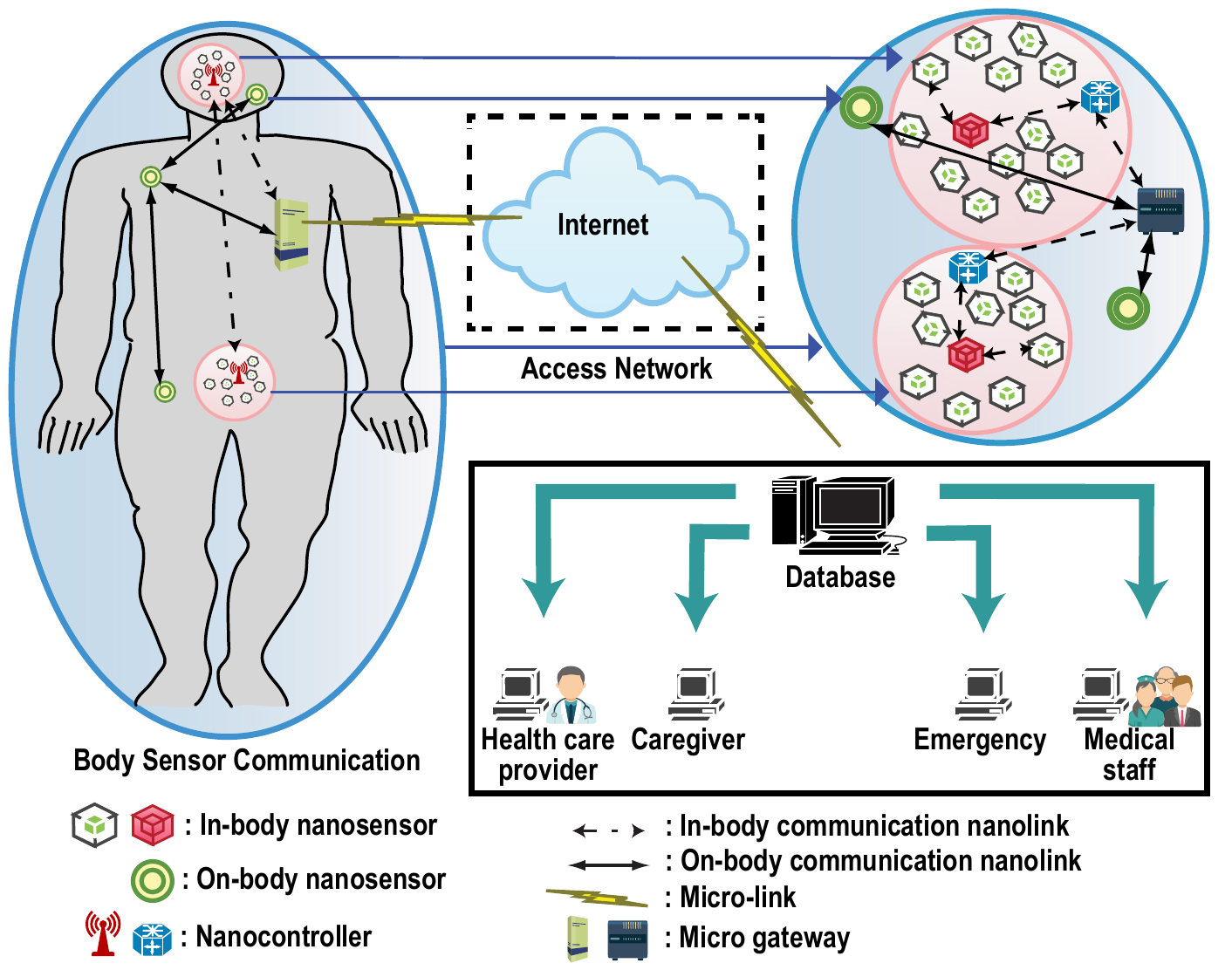}
 \caption{\textbf{A possible communication architecture}. The network is composed of nanonodes, nanocontrollers, and a micro gateway. The in-body and on-body nanosensors are placed to accumulate data from inside and outside of the human body, respectively. Nanocontrollers collect data from the nanosensors and forward them to the micro gateway, which is responsible for storing the data in a remote database using the Internet, and these data can be accessed by the relevant entities.}
      \label{fig1}
\end{figure*}

\subsection{Terahertz Channel Modeling}
\label{subsec-thzchanmodel}
Channel modeling is another important issue in EM communications at the nanoscale over the THz band. Jornet and Akyildiz proposed a physical channel model and used it to investigate signal path loss, molecular absorption, and channel capacity of the network \cite{23jornet2011}. Radiative transfer theory was used in computing total path loss of a signal traveling short distances to assess molecular absorption. Different power allocation patterns were proposed for the THz band including a scheme based on the transmission of femtosecond-long pulses; using these power allocation schemes, the performance of the THz band was evaluated for channel capacity. The propagation model proposed by Piro et al. characterized the performance of THz communication in human tissue considering the attenuation of EM waves in human skin tissues \cite{24piro2015}. They deduced channel capacity and communication ranges for different physical transmission settings. Path loss and molecular absorption noise temperature were also obtained using optical parameters of human skin tissues and were verified through extensive experimental tests. In addition, Javed and Naqvi proposed SimpleNano, a channel model for WNSNs that is capable of transmitting in the THz range \cite{25javed2013} where a log-distance path loss model together with random attenuation was approximated in mediums such as the human body.

\section{System Model}
\label{sec-systmodel}
\subsection{Network Model}
\label{subsec-netwmodel}
In modeling the network, it is assumed that a network of nanosensor nodes, constrained by energy, is placed into a human body to construct a body sensor network (BSN). A possible system architecture is shown in Fig.  \ref{fig1}.
The communication is expected to be from inside to outside of the body through nanosensor nodes. Nanocontrollers (NC) are deployed for in-body communication, which aggregate information coming from in-body nanosensors. The micro gateway, a wireless interface connecting the subnetwork of nanosensors to the Internet, collects data and forwards them to the relevant entity (e.g., healthcare provider, medical staff, etc.) using the Internet and/or intranet.

\subsection{Proposed Approach}
\label{subsec-proposedapproach}
\begin{figure}[htb!]
  \centering
    \includegraphics[scale=1]{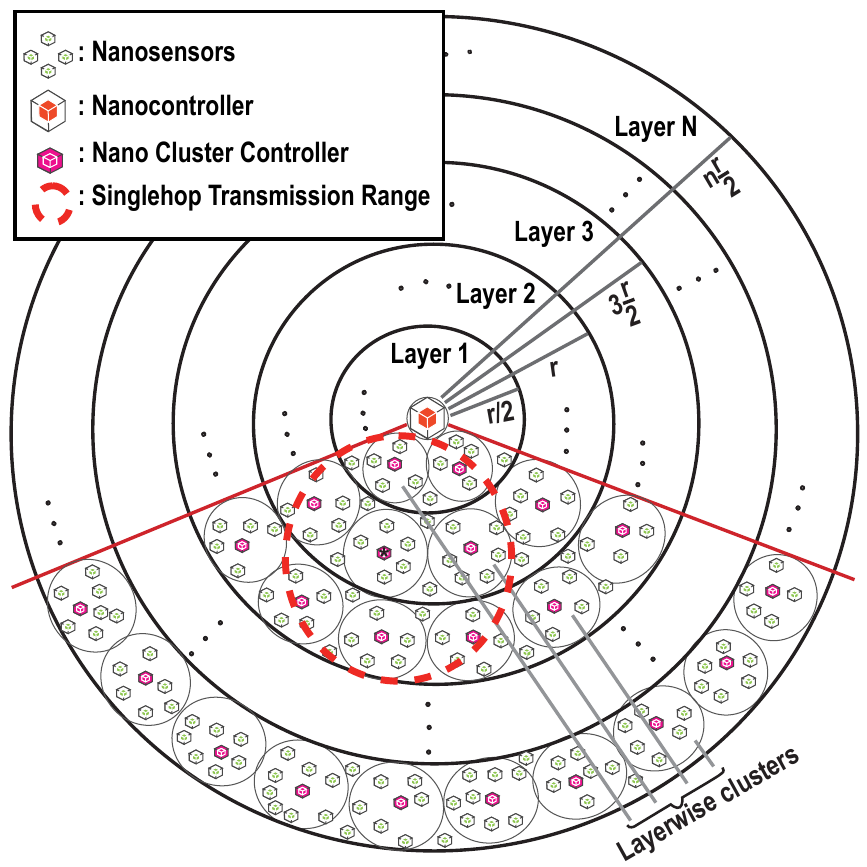}    
    \caption{\textbf{Possible architecture of proposed scheme}. Nanosensors are deployed within their associated concentric layers of radius $n\frac{r}{2}$, where $n$ is the number of layers and $r$ is the transmission range of each nanosensor. The singlehop transmission range for a nano cluster controller, marked with *, is indicated by the orange dashed-circle. The nanocontroller is assumed to be placed at the center of all layers. Nanosensors form clusters under each layer follow the proposed algorithm \ref{algor2-ncca}.}
    \label{fig2-golla}
\end{figure}

The proposed approach (see Fig. \ref{fig2-golla}) uses a multi-layered topology based on the distance of deployed nanosensors from NC. To increase network stability and achieve high energy efficiency, each layer is divided into multiple nano-clusters and re-clustering is avoided. Nano cluster controllers (NCCs) from each layer are elected based on the residual energy of nanosensors. NCCs collect data from nano-cluster members (NCMs) and forward aggregated data towards NC. Clusters of one layer are capable of communicating with clusters of the immediate upper or lower layers. It is assumed that nanosensors are capable of controlling power dynamically. NCC conducts intra-cluster communication using low power transmission range and inter-cluster communication using high power transmission range.

A major problem in WNSN communication is the high probability of link failure due to low transmission power and high transmission losses (e.g., path loss and shadowing). This problem can be addressed successfully using cooperative communication among NCCs (i.e., amplify-and forward, and decode-and-forward \cite{nosratinia2004,chen2014}) and information fusion at the fusion node \cite{kampis2015}. The role of the fusion node is executed by either the destination NCC or the NC where the information fusion occurs. 

\subsection{Network Configuration}
\label{subsec-netwconfig}
The network is organized in layers to help the nanosensors discover the layers they belong to. Assuming the sensing field to be circular, at first the whole network is divided into several virtual layers with respect to the NC. The radius of each layer is computed as $n\frac{r}{2}$ where $n$ is the number of layers and $r$ is the transmission range of each nanosensor. Layers are divided in a manner that ensures the transmission range of each nanosensor of a particular layer spans its adjacent layers. Upon receiving a broadcasted message from NC, all the nanosensors calculate their distances from NC, compare their distances with the layer radius, and register themselves to appropriate layers. The mechanism of Nanosensor Distribution over Layers (NDL) is presented in algorithm \ref{algor1-ndl}.

\begin{algorithm}[hbt!] %or another one check

 \caption{NDL Algorithm}% for Nanosensor Distribution over Layers }
     \SetAlgoLined

     Notations:\\
  $k$ = Number of deployed nanosensors\\
  $d_k$ = Distance of nanosensor $k$ with respect to NC\\
  $RSSI$ =Received Signal Strength Indicator  \\
 $L_k$ = Layer number to which node $k$ is assigned\\
  $r$ = Transmission range of each nanosensor \\

     \KwData{$r$, $k$}
     \KwResult{$L_k$}
     NC broadcasts ``Hello packets'' to all $k$\;

    \ForEach{$k$}
    {
    Calculate $d_k$ with respect to NC based on received $RSSI$\;
    $L_k \longleftarrow \textrm{floor}(2 \frac{d_k}{r})$ \;
    $k$ registers itself into layer $L_k$\;

    }
		\label{algor1-ndl}
\end{algorithm}
%\linespread{1.5}

\subsection{Nano Cluster Composition}
\label{subsec-nanoculstcomp}
After all nanosensors are assigned to their respective layers, the initial NCCs are automatically elected based on their residual energy. This election process involves, at the initial round, the NCC candidates competing for the ability to upgrade to NCC by multicasting their weight, denoted by $\zeta_k$ and calculated using eq. \eqref{ef0}, to neighboring candidates.
\begin{equation}
%\small{
\label{ef0}
  \zeta_k = \frac{E_{residual_k}}{E_{max}}.
\end{equation}
where,
$E_{residual_k}$ is the residual energy of nanosensor $k$, and $E_{max}$ is the maximum energy.

If a given nanosensor does not find another nanosensor with larger $\zeta_k$, it declares itself as NCC. After NCCs have been elected, cluster formation begins. The new NCCs multicast short range advertisement messages to all non-clustered nanonodes within the same layer. Each of the non-member nanosensors having a higher received signal strength indicator (RSSI) sends a join request to its respective NCC and registers itself as an NCM. This process continues until clusters are formed and all nodes are assigned to their respective NCCs. To reduce clustering overhead, clustering is done only once for the whole network. Only the NCC will be changed after each transmission using a round robin process. The NCCA is presented in algorithm \ref{algor2-ncca}.

 \begin{algorithm}[bht!] %or another one check

 \caption{NCCA for cluster formation}
     \SetAlgoLined

     Notations:\\
  $k$ = Number of deployed nanosensors\\
  $\zeta_k$ = Normalized residual energy of sensor $k$ \\
  $k_{max(\zeta)}$ = Nanosensor with maximum $\zeta$\\
 $GN$ = General Node\\

     \KwData{$L_k, E_{residual},  E_{max}$}
     \KwResult{$Clusters$}

    \ForEach{$k$}
    {
    Calculate $\zeta_k$ using eq. \eqref{ef0}\;
            Multicast $\zeta_k$ within associated $L_k$\;
                {\eIf{$k_{max(\zeta)}$}{
                 $Status \longleftarrow NCC$\;
                 %break\;
                    }
                    {$Status \longleftarrow GN$\;}
                    }
    }

     Elected $NCC$ multicasts short range advert message to $GN$ within associated $L_k$\;

 \ForEach{$GN$}
    {
     Check received $RSSI$ from all NCCs within range\;
    Send join request to NCC as NCM for which $RSSI$ is maximum and get registered\;
 }
\label{algor2-ncca}
\end{algorithm}

\begin{figure*}[!bth]
    \centering
    \includegraphics[scale=1]{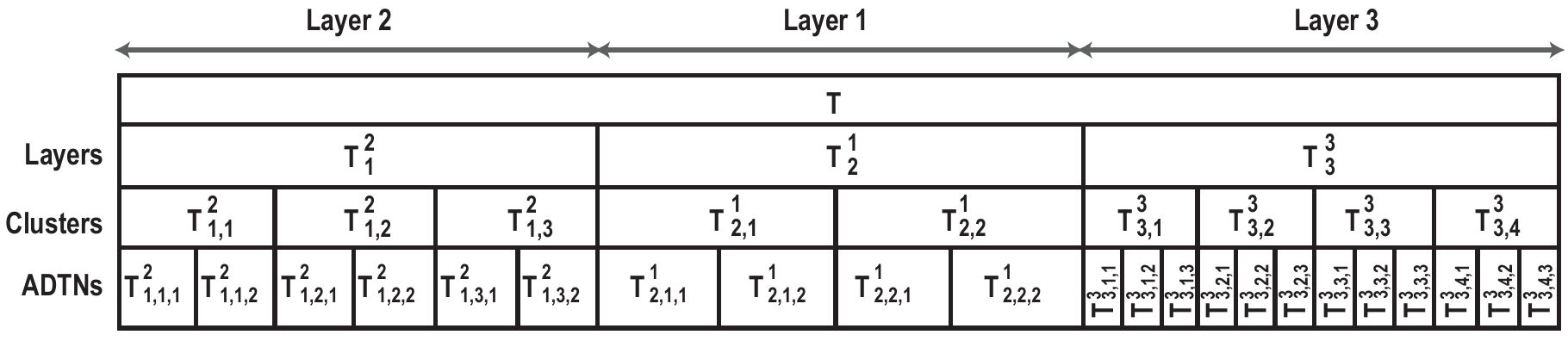}
    \caption{\textbf{Time scheduling mechanism}. It is assumed that there are three homocentric layers in the network. Transmission cycle T is divided into three variable length time slots ($T_1^2$, $T_2^1$ and $T_3^3$) based on the data requirements of layers. Each timeslot is further distributed among clusters and ADTNs based on the value of $\alpha$ (eq. \eqref{ef1}) by the nanocontroller.}
    \label{fig3-time}
\end{figure*}

\subsection{Communication Mechanism}
\label{subsec-commmechanism}
The sensed data are aggregated and forwarded to NC in multihop fashion using a novel TDMA based dynamic scheduling scheme. In our scenario, to collect information from the whole BSN, the NC periodically seeks data transmission request packets from all the nanosensors. In response, each Active Data Transmission Node (ADTN, i.e., an NCM with data to transmit) sends transmission request packets whose structure is shown in table \ref{table:3}.

\begin{table}[!bt]
\renewcommand{\arraystretch}{1.7}
\caption{Structure of transmission message request forwarded to NC}

\begin{center}
    \begin{tabular}{|l|l|l|l|l|l|}

    \hline
    NCM ID & NC ID &$d_k$ & $L\#$ & $E_{residual}$ & $D_{Amnt}$ \\ \hline
    \end{tabular}
\end{center}
\label{table:3}
\end{table}
Each of these packets consists of information about the ADTN (NCM ID) and NC (NC ID), ADTN’s distance from NC ($d_k$), ADTN's layer information ($L\#$), amount of data to be sent ($D_{Amnt}$), and the residual energy ($E_{residual}$). In turn, the NC will prioritize data to be received, estimate total time considering propagation delay for a complete transmission and specify packet transmission order for the ADTNs. Transmission timeline scheduling is done in the following three phases:

\noindent
\textbf{Phase 1:}
The NC allocates a variable-length transmission time slot for each layer. The slot length depends on the transmission parameters, i.e., total amount of transmittable data of the layer, total time required for the data to be transmitted, and priority of the data. The NC calculates the layer ordering weight - $\alpha$ - employing eq. \eqref{ef1} and determines the layer order of data transmission. 
\begin{equation}
\label{ef1}
  \alpha = \{L\times(t+\Gamma)T^{-1}\times(\sum_{\iota\in m}\sum_{\nu\in \kappa}M_{\iota,\nu}^L)\}\times P,
\end{equation}
where,
$\alpha$ is the layer ordering weight; $L$, $\iota$, $\nu$ are the IDs of layer, cluster, and node respectively; $t$, $T$ are the transfer time per data unit for a layer and transmission cycle completion time; $\Gamma$ is the propagation delay; 
$m$ is the number of clusters in a layer; $\kappa$ is the number of ADTNs in a cluster; $P$ is the priority of data to be sent; and $M$ is the amount of data to be transmitted.

\noindent
\textbf{Phase 2:}
The NC then divides the total allocated time for a particular layer among its clusters based on the amount of data the cluster needs to transmit. This will prevent TDMA intra-cluster transmission from colliding with other cluster transmissions from the same layer.

\noindent
\textbf{Phase 3:}
Based on the assigned time slots for each cluster, each ADTN under it is allocated a new sub time slot using the mechanism shown in algorithm \ref{algor3}.

 \begin{algorithm}[bht!] %or another one check

 \caption{Algorithm for time slot allocation}
     \SetAlgoLined

     Notations:\\
  $\alpha$ = Weight upon which layers are ordered\\
  $LL$ = List of prioritized order of layers\\
  $p$= $1,2,\cdots n$; 
  $T_p^L$ = Time slot for layer $L$\\
  $T_{p,\iota}^L$ = Time slot for Cluster $\iota$ of layer $L$\\
  $T_{p,\iota,\nu}^L$ = Time slot for ADTN $\nu$ of cluster $\iota$ of layer $L$\\
     \KwData{$Transmission \:request \:message $}
     \KwResult{Variable \:length \:time\: slots \:for\: all\: ADTNs}
NC relays message to NCMs requesting transmission\;
ADTNs send transmission request message to NC\;
Calculate $\alpha$ using eq. \eqref{ef1} for each layer\;
Calculate total time $T$ for a complete cycle\;
Sort $\alpha$ into descending order \;
Set priority of layers according to the order\;
For each layer create prioritized layer list $LL$\;
    \ForEach{$LL$}
    {
    Calculate layer time $T_p^L$\;

 \ForEach{$T_p^L$}
    {
     Calculate each cluster time $T_{p,\iota}^L$\;
   \ForEach{$T_{p,\iota}^L$}
    {
     Calculate ADTN time $T_{p,\iota,\nu}^L$\;
     Allocate time slots accordingly\;

 }
 }
 }
\label{algor3}
\end{algorithm}

Fig. \ref{fig3-time} shows an exemplary time scheduling procedure with a 3 layer network which requires $T$ time to complete a transmission cycle. Using $\alpha$ value (calculated by eq. \eqref{ef1}) the NC prioritized the transmitting layers as layer 2, layer 1, and layer 3 during $T_1^2, T_2^1, T_3^3$ time slots. Each layer 1, 2, and 3 has 2, 3, and 4 clusters with 2, 2, and 3 ADTNs to transmit. The time slot scheduling is prioritized based on the amount of data each cluster needs to transmit from a layer. In this example, layer 2 ($T_1^2$) is scheduled in 3 time slots each for a cluster ($T_{1,1}^2, T_{1,2}^2, T_{1,3}^2$) which are then distributed among ADTNs (i.e., $T_{1,1}^2$ is divided into $ T_{1,1,1}^2$, $T_{1,1,2}^2$ and allocated to the $2$ transmitting ADTNs). This procedure is repeated for all transmitting clusters and layers. This proposed time scheduling scheme utilizes optimally the large bandwidth provided by the THz band and ensures collision avoidance both at NCCs and NC.

\subsection{Data Transmission}
\label{subsec-datatrans}
At the beginning, all nanosensors - except NCCs - remain in harvesting mode. After the time allocation, NC sends a wake-up preamble to activate the layer with the highest priority, in turn, activating all the ADTNs of that layer for transmission.

In the intra-cluster period (see algorithm \ref{algor4}), ADTNs forward their data to respective NCCs, either directly (i.e., ADTN to NCC) or via another NCM, depending on energy per bit consumption to transmit. Considering $i$ as transmitting ADTN, $j$ as NCM and $z$ as NCC, and denoting the average energy required to transmit a bit by $E(d_{iz})$ using $i-z$ link, $E(d_{ij})$ using $i-j$ link, and $E(d_{jz})$ using $j-z$ links, the data is transmitted directly from $i$ to NCC if,
\begin{math}
\label{ee2}
  \{E(d_{ij}) + E(d_{jz})\}> E(d_{iz}),
\end{math}
otherwise, an intermediate hop is used and the time slot of ADTN is divided equally for ADTN-NCM and NCM-NCC communication.

The inter-cluster--inter-layer period starts (see algorithm \ref{algor4}). The NCCs aggregate data and forward them to NCCs of an adjacent lower layer. After forwarding data, NCCs call for new NCC election using round-robin scheduling. New NCCs then multicast their new status to other NCMs. Previous NCCs give their information list to new NCCs and go to harvesting mode. All nodes except the new NCC start harvesting and the data transmission mode is paused until another wake-up message arrives. 

	 \begin{algorithm}[t!] %or another one check
		
		\caption{Algorithm for the data transmission process}
		\SetAlgoLined
		
		\KwData{$LL$}
		\KwResult{A Cycle of data transmission}
		
		\ForEach{$LL$}
		{
			NC sends wake up preamble\;
			
			\While{$intra-cluster \:period$}
			{ADTNs send data to respective NCC using $T_{p,\iota,\nu}^L$ time slots\;
				NCC aggregates data \;
			}
			
			\While{$inter -\:cluster--inter- \:layer\: period$}
			{
				{	\eIf{$L=1 \:\&\&\: NCC_{\iota}^L$ receives multiple links}
				{$NCC_{\iota}^L$ performs fusion \;
			$NCC_{\iota}^L$ transmits fused data to NC\;	
			}
				{$NCC_{\iota}^L$ performs DaF scheme to transmit data to lower layer $NCC_{\iota}^{L-1}$\;
			}}
				
			New NCC is elected using round robin fashion\;
				New NCC multicasts update to other NCMs\;
				Previous NCC transfers its information list to new NCC\;
				All ADTNs and previous NCC go to harvesting mode except the new NCC until another wake up preamble reaches\;
				
			}

		}
		\label{algor4}
	\end{algorithm}

\begin{figure*}[htbp!]
		\centering
		\includegraphics[scale=1]{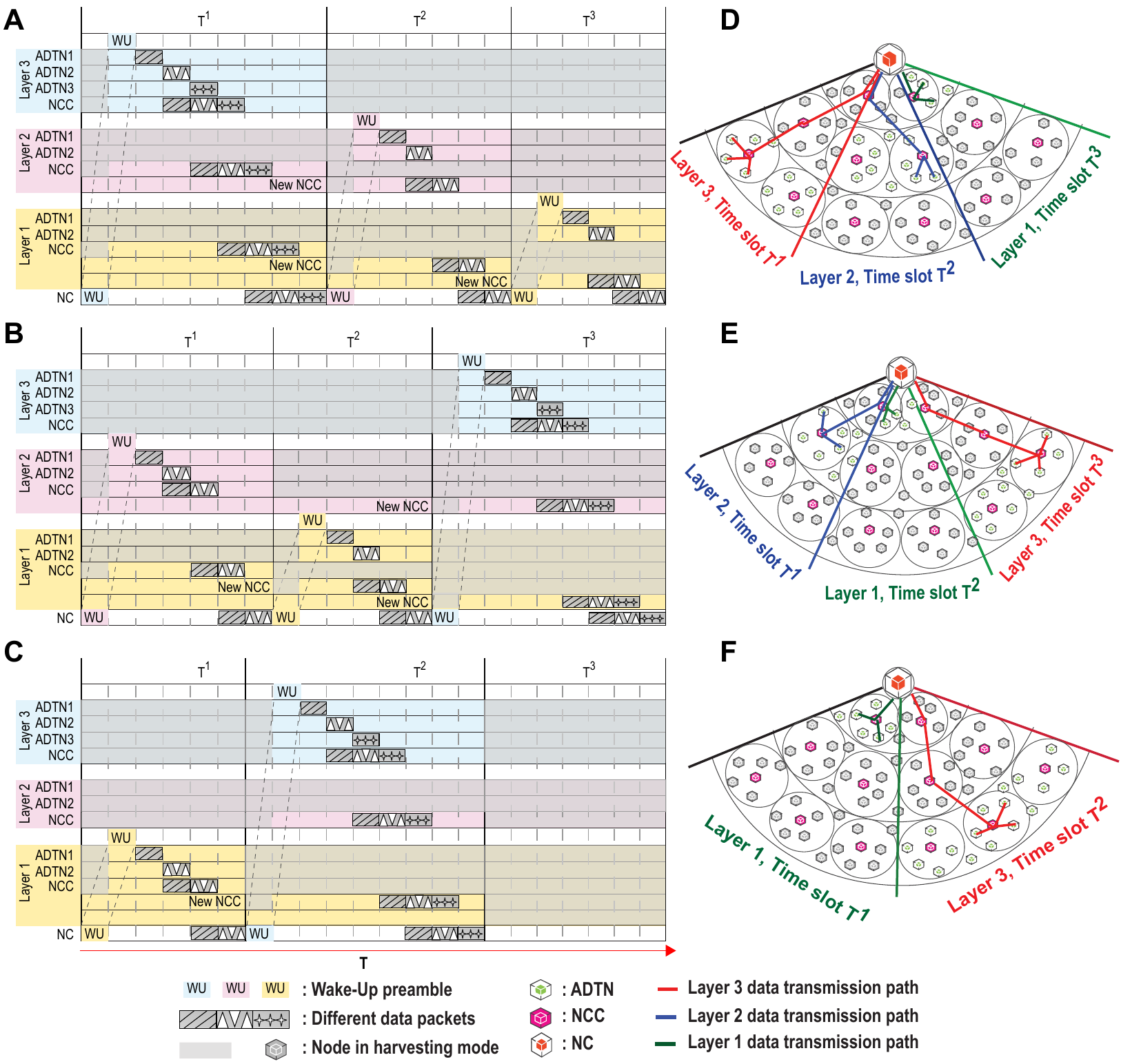}
		\caption{\textbf{Possible cases of data transmission.} Timeslot allocation (\textbf{A}-\textbf{C}) and data transmission processes (\textbf{D}-\textbf{F}) are shown for three different cases. As per the data transmission mechanism, NC calculates the $\alpha$ value (see eq. \eqref{ef1}) to prioritize different layers' data dispatching and assigns them the necessary timeslots to relay. Case 1 (\textbf{A},\textbf{D}): layers are prioritized as layer 3, layer 2, and layer 1 to transmit during $T_1$, $T_2$, and $T_3$ timeslots. Case 2 (\textbf{B},\textbf{E}): the priority of the layers are layer 2, layer 1, and layer 3 and they transmit during $T_1$, $T_2$, and $T_3$ timeslots, respectively. Case 3 (\textbf{C}, \textbf{F}): only two layers transmit in the order layer 1 and layer 3 during $T_1$ and $T_2$ timeslots.}
	\label{fig4}
\end{figure*}

\begin{figure}[htbp!]
		 \centering
		\includegraphics[scale=1]{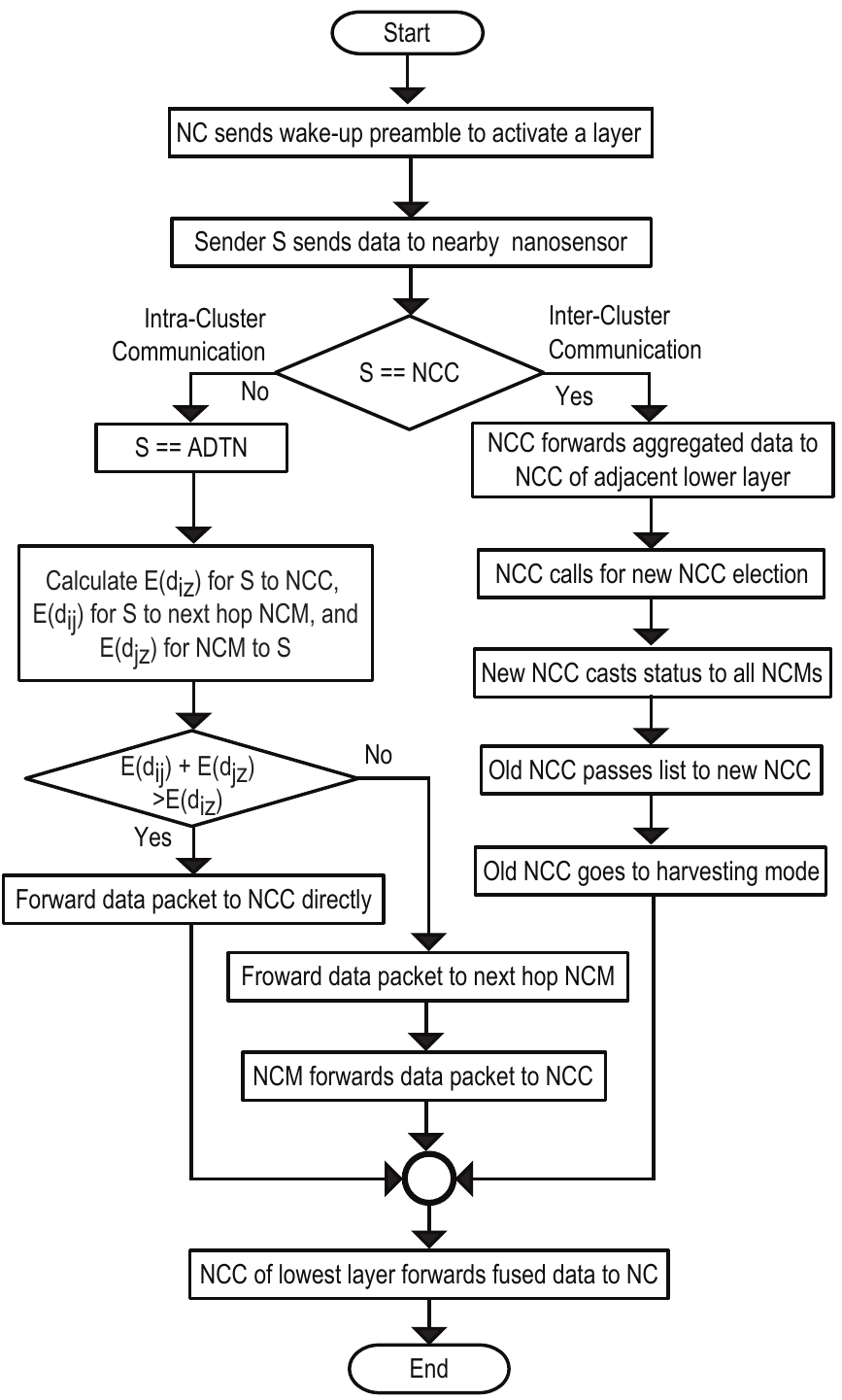}
	\caption{\textbf{A flowchart of the communication framework.} The communication mechanism consists of inter-cluster and intra-cluster communication. After a layer is activated, data transmission begins. In intra-cluster communication, all ADTNs send their data packet to the NCC within their associated cluster. NCC fuses data and conducts inter-cluster communication towards the NC. Once a transmission is completed by an NCC, it calls for new NCC election within the same cluster for conducting succeeding transmission. After a new NCC is elected, previous NCCs with retired ADTNs harvest until another wake-up call arrives from the NC.}
	\label{fig7-flow}
\end{figure}

Fig. \ref{fig4} shows three possible transmission scenarios. In the first scenario (A and D) where the priority order of the 3 layers are set as layer 3, layer 2, and layer 1. To initiate the transmission, the NC first sends a wake-up preamble to layer 3. The ADTNs belonging to layer 3 are activated in response to the preamble and set to data transmission mode while the remaining ADTNs stay in harvesting mode. In addition to the ADTNs of layer 3, the NCCs of all layers are activated during the transmission process. Once the data transmission is completed, new NCCs are elected and all other ADTNs along with previous NCCs return to harvesting mode.
In the second scenario (B and E) the transmission layer order is changed to layer 2, layer 1, and layer 3. Here, time slots are allocated for layer 2 first, and the wake-up preamble is sent to layer 2 to complete the transmission which is then followed by layer 1 and finally by layer 3.
In the third scenario (C and F), only two layers transmit data in the order layer 1 and layer 3. As layer 2 does not transmit, no time slot has been allocated for it. In the first time slots upon arrival of the wake-up preamble layer 1 transmits and then layer 3 transmits in the second time slot.
The generic flow chart of the data transmission process is shown in Fig. \ref{fig7-flow}.

\begin{figure}[htbp!]
	\centering
	\includegraphics[clip]{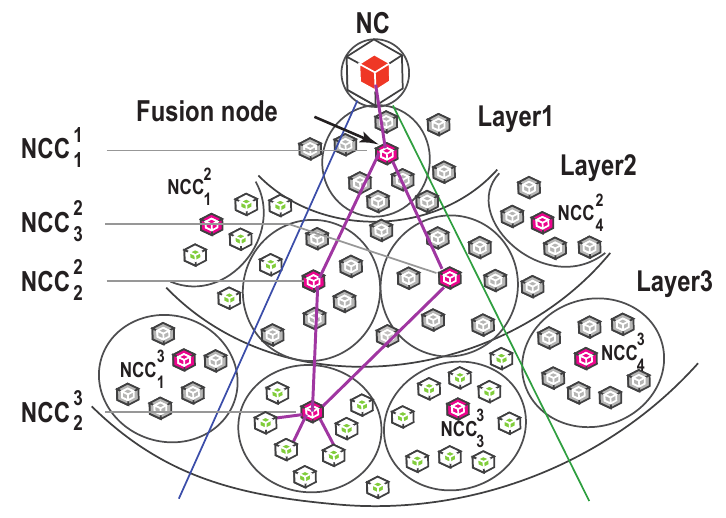}
	\caption{\textbf{Data fusion using cooperative communication.} A conceptual overview of a fusion mechanism is shown. All intermediate NCCs between the source and fusion node performs DaF to transmit data; the last hop NCC performs fusion to eliminate data redundancy.}
	\label{fuse}
\end{figure}

A transmission with decode-and-forward (DaF) strategy involving information fusion is shown in Fig. \ref{fuse}. In DaF, the cooperative NCCs ($NCC^L_\iota$; where $L$ is the Layer and $\iota$ is the NCC number) successfully receive, decode and forward the transmission to the adjacent lower layer ($NCC^{L-1}_\iota$) using a distributed space-time code \cite{alamouti1998}. Since the intermediate NCCs decode the message successfully and forward it to the next layer's NCC(s), no outage is expected to occur during these hops. This fusion helps to reduce information outage and data redundancy.

\subsection{Channel Behavior Model}
\label{subsec-chanbehavmodel}
An EM wave in THz band experiences several attenuations (in particular absorption and spreading losses) while propagating through human tissues. Both link distance $d_{i,j}$ (for $i-j$ link) and signaling frequency $f$ influence these attenuation.

Molecular absorption causes attenuation or absorption loss depending on the type of molecular composition constituting the medium. The human body is the transmission medium when placed nanosensors constitute the BSN. In the human body, water molecules outnumber other molecules, thus, the value of the molecular absorption coefficient is given by the weighted average of different absorption coefficients of water molecules present in different body parts \cite{25javed2013}.

In the THz band BSN, the wave propagation range through human tissues is constrained by a large spreading loss \cite{24piro2015} and shadowing \cite{brizzi2013}. Shadowing, caused by body part movement, increases the blocking probability because of broken line-of-sight links between the transmitter and receiver \cite{brizzi2013}.

If $P_t(i)$ is the transmitted power at the source node $i$, then the total received power $(P_{total}(i,j))$ at the destination $j$ considering the molecular absorption, spreading, and shadowing effect is given by eq. \eqref{eq6} \cite{afsana2015Outage}.
\begin{equation}
\label{eq6}
	P_{total}(i,j)=\overbrace{\Big(P_t(i) G d_{i,j}^{-\eta} 10^{\frac{\xi_{i,j}}{10}}\Big)}^{ \text{RPS}}\overbrace{\Big(e^{-K(f)d_{i,j}}\Big)}^{\text{AL}}\overbrace{\Big(\frac{4\pi fd_{i,j}}{c}\Big)^{-2}}^{\text{SL}}
\end{equation}
where $G$ is the gain constant, $\eta$ is the path loss exponent, $\xi_{i,j}$ is a normal distributed random variable with zero mean and $\sigma$ standard deviation, $K(f)$ is the molecular absorption coefficient, and $c$ is the speed of light. The RPS component is the `received power due to shadowing', the AL is `absorption loss', and the SL is `spreading loss'.

The interference power, $P_I(z,j)$ for the $z-j$ link with $z \in \textbf{\textit{k}}=\{\textrm{interfering links}\}$, is computed by eq. \eqref{eq7}.
\begin{equation}\label{eq7}
	P_I(z,j) = \sum_{z \in \textbf{\textit{k}}}\frac{ P_t(z) G 10^{\frac{\xi_{z,j}}{10}}c^2e^{-K(f)d_{z,j}}}{16\pi^2f^2d_{z,j}^{\eta+2}}.
\end{equation}

The $j$th hop's SINR ($\gamma_j$) can be computed by fusing Equations \eqref{eq6} and \eqref{eq7} and considering additive white Gaussian noise power at destination $j$ ($N_j$) using eq. \eqref{eq8}  \cite{afsana2015Energy}.

\begin{equation}\label{eq8}
	\gamma_j = \frac {P_{total}(i,j)}{P_I{(z,j)}+N_j}.
\end{equation}

Because path loss due to shadowing at the molecular level follows lognormal distribution (LND), $\gamma_j$ also follows LND under lognormal shadowing and is approximated by a lognormal random variable with standard deviation ($\sigma_j$) and mean ($\varphi_j$) calculated by the Fenton-Wilkinson method \cite{32kelif2010}.

Outage ($P_{out}$) is given by eq. \eqref{eq111} with $Q[\cdot]$ error function when  $\gamma_j$ falls below a threshold ($\gamma_{th}$) \cite{afsana2015Energy}.
\begin{eqnarray}
\label{eq111}
  P_{out}&=&Pr(\gamma_j < \gamma_{th})\nonumber\\
	&\approx& Q\left[\frac{10\log_{10}{\gamma_{th}-\varphi_j}}{\sigma_j}\right].
\end{eqnarray}

The total channel capacity ($C$) is obtained by summing individual capacities of the sub-channels using eq. \eqref{ec} \cite{goldsmith2005}.
\begin{equation}\label{ec}
	C = \sum_i B_i \log_2(1+\gamma_i).
\end{equation}
where $B_i$ is the bandwidth of each sub-channel $i$.

In the DaF strategy \cite{nosratinia2004}, the outage is calculated using eq. \eqref{eq:poutDaF}. 

\begin{equation} 
\label{eq:poutDaF}
	P_{out}=\prod_{y} Pr({\gamma_{y}<2^{nC/B}-1})
\end{equation}
where $y$ is the number of multipaths between the source and the fusion node, $\gamma_{y}$ is the $y$th link's SINR from the source node to the fusion node, and $n$ is the number of layers.

The outage capacity probability, $C(P_{out})$, determines the constant data rate in non-outage fading states using eq. \eqref{ec1} and denotes the actual capacity of a transmitting channel \cite{afsana2015Outage}.

\begin{equation}\label{ec1}
C(P_{out}) =  B \log_2(1+\gamma_j)(1-P_{out}).
\end{equation}

\begin{figure*}[htb!]
	\centering
		\includegraphics[width=1\linewidth]{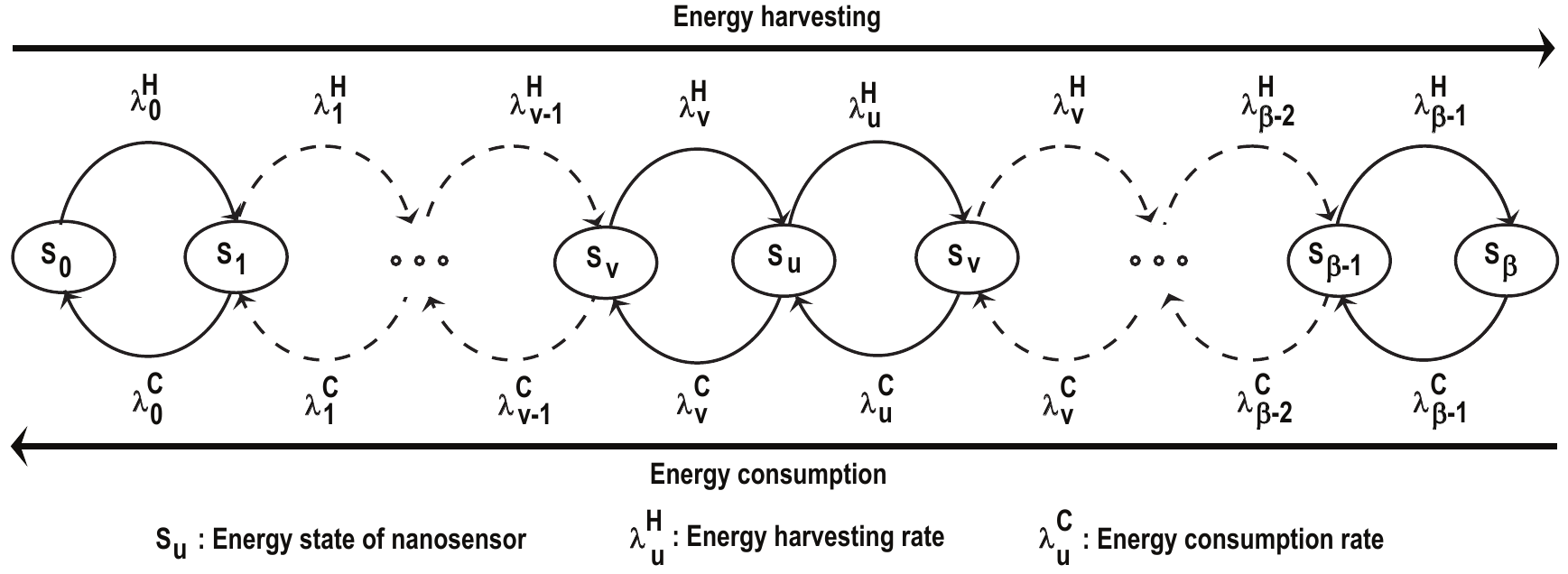}
	\caption{\textbf{A Partial Markov decision process (MDP) for a nanosensor energy model}. The energy harvesting and consumption process is modeled by representing states and transitions using nano-stationary continuous-time Markov process, $\varepsilon(t)$.}
	\label{markov}
\end{figure*}

\subsection{Energy Model of the Nanosensor}
\label{subsec-energymodel}
The nanosensor energy is modeled using nano-stationary continuous-time Markov process, $\varepsilon(t)$, which describes the evolution of the energy states in time $t$ \cite{26jornet2012}. $S_u$ is the energy state of a nanosensor; at state $S_0$ it has the minimum energy ($E_{min}$) to operate, and at state $S_{\beta}$ it has full energy. Here, $\beta$ is the number of cycles needed to charge the nano power source (e.g., Piezoelectric nanogenerator as in \cite{26jornet2012}, Pyroelectric nanogenerators as in \cite{yang2012}, etc.) up to an energy value $E$. The Markov chain of the model without self transitions is shown in Fig.  \ref{markov} and is defined by its transition rate matrix $\Omega(t)$ calculated by eq. \eqref{mareq}. 
Each element of the matrix $\omega_{S_uS_v}$ refers to the transition rate from state $S_u$ to state $S_v$ and is defined in eq. \eqref{eq:dd}.

\begin{equation} \label{eq:dd}
\omega_{S_uS_v} = \left\{
\begin{array}{llll}
\lambda_u^H & S_0\leq S_u\leq S_{\beta-1} ,\; S_v=S_{u+1};\\
\lambda_u^C & S_1\leq S_u\leq S_{\beta} ,\; S_v=S_{u-1};\\
-\omega_{S_uS_w} & S_u = S_v;\\
0 & \mbox{Otherwise.}
\end{array}
\right.
\end{equation}

where $\lambda_u^H$ is the energy harvesting rate, $\lambda_u^C$ is the energy consumption rate, and $\omega_{S_uS_w}$ is defined as $\sum_{S_w=1,\;S_w\neq S_u}^{S_{\beta}}\omega_{S_uS_w}$.

\begin{multline}
\label{mareq}
	\Omega\{t\} = \\
	\resizebox{\linewidth}{!}{\bordermatrix{
			&	S_0 & S_1			 & \cdots	& S_{\beta-1} & S_{\beta}  \cr
			S_0 & -\lambda_0^H   & \lambda_0^H   & \cdots 	& 0 & 0 \cr
			S_1 &\lambda_0^C     & -(\lambda_1^H+\lambda_0^C)& \cdots & 0 & 0 \cr
		\vdots & \vdots & \vdots& \ddots&\vdots& \vdots\cr
			S_{\beta-1} &0               & 0             & \cdots   &-(\lambda_{\beta-1}^H+ \lambda_{\beta-2}^C) & \lambda_{\beta-1}^H  \cr
			S_{\beta}  &0                &0              &  \cdots       & \lambda_{\beta-1}^C & -\lambda_{\beta-1}^C \cr\\
	}}
\end{multline}

For a power source of $\beta$ cycles with cycle length $\tau$, $V_{nps}$ is computed using the expressions shown in eq. \eqref{ea1}.

\begin{eqnarray}
 S_0 \leftrightarrow S_1: 1\,\mbox{cy.} : V_{nps}(1) &=& V_g \big(1-e^{-\tau\diagup R_g C_{nps}}\big)\nonumber   \\
 S_0 \leftrightarrow S_1 \leftrightarrow S_2: 2\, \mbox{cy.} : V_{nps}(2) &=& V_g \big(1-e^{-2\tau\diagup R_g C_{nps}}\big)\nonumber   \\
 &\vdots& \nonumber \\
 S_0 \leftrightarrow \cdots \leftrightarrow S_{\beta}: \beta\, \mbox{cy.} : V_{nps}(\beta) &=& V_g \big(1-e^{-\beta \tau\diagup R_g C_{nps}}\big)\nonumber \\
 &=& V_g \big(1-e^{-\beta \Delta \Omega\diagup V_g C_{nps}}\big).\nonumber\\
\label{ea1}
\end{eqnarray}
where $V_g$ is the generator voltage, $R_g$ is the resistance of the nano power source, and $\Delta \Omega$ is the harvested charge per cycle (cy.).

The maximum energy stored in the nano power source, $E_{nps-max}$, is estimated using eq. \eqref{eqenps-max}.
\begin{eqnarray}
\label{eqenps-max}
 E_{nps-max} &=& \max{\big\{E_{nps}(\beta)}\big\}\nonumber   \\
  &=&  \max{\big\{\frac{1}{2}C_{nps}\big(V_{nps}(\beta)\big)^2\big\}}.
\end{eqnarray}
where $E_{nps}$ is the energy stored in the nano power 
source  \cite{26jornet2012}, $C_{nps}$ is 
the total capacitance of the nano power source, and 
$V_{nps}$ is the voltage of the charging power source. 

When the value of $\beta$ approaches $\infty$, the voltage of the charging power source ($V_{nps}$) becomes equal to the voltage of the generator ($V_g$). In this case, the maximum energy ($E_{nps-max}$) is given by eq. \eqref{enpsmax-binf}.

\begin{equation}
	E_{nps-max} = \frac{1}{2} C_{nps} V_g^2.
\label{enpsmax-binf}
\end{equation}

The number of cycles ($\beta$) needed to charge the power source to an energy value ($E$) is derived as in eq. \eqref{en2}.
\begin{equation}\label{en2}
  \beta(E) = \Bigg\lceil - \frac{V_g C_{nps}}{\Delta \Omega} \ln \Big( 1-\sqrt{\frac{E}{E_{nps-max}}}\,\Big)\,\,\Bigg\rceil
\end{equation}

If the energy of the power source is increased by $\Delta E$, then the energy harvesting rate, $\lambda^H$, of the nano power source is computed using eq. \eqref{en3}.
\begin{equation}\label{en3}
  \lambda^H(E_{nps}, \Delta E) = \frac{\beta \Delta E}{\tau}\frac{1}{\beta(E_{nps}+ \Delta E)- \beta(E_{nps})}.
\end{equation}

Based on eq. \eqref{en3}, the harvesting rate for transition from energy state $S_u$ to state $S_{u+1}$, $0 \leq u \leq \beta-1$, is given by eq. \eqref{eq:energyharv}; and the energy consumption rate $\lambda_u^C$ is given by eq. \eqref{eq:energyconsump}.
\begin{equation}
\lambda_u^H = \frac{\lambda(E_u,E_{tx})}{E_{tx}}.
\label{eq:energyharv}
\end{equation}
where $E_{tx}$ is the energy required to transmit a packet, and $E_u$ is the energy at state $u$ and is given by $E_{min} + u \times E_{tx}$.

%The energy consumption rate $\lambda_u^C$ is given by eq. \eqref{eq:energyconsump}.
\begin{equation}
\label{eq:energyconsump}
\lambda_u^C = \frac{\lambda_x^C}{E_{tx}}.
\end{equation}

Given the energy required for multi and singlehop transmission, the energy saving probability for multihop transmission, $P_{ES}$, is defined by eq. \eqref{eq:pes}.
\begin{eqnarray}
	P_{ES} &=& Pr\{E_{b_{SR}}+E_{b_{RD}} < E_{b_{SD}}\}\nonumber \\
				 &=& 1-e^{-\theta(E_{b_{SR}}+E_{b_{RD}})}
\label{eq:pes}
\end{eqnarray}
where $E_{b_{--}}$ is the average bit energy required for transmission; $SR$ is the transmission from source (ADTN/ NCC) to a relay node (NCC), $RD$ is the transmission from the relay node to destination (NC), $SD$ is the direct transmission from source to destination; and $\theta$ is the rate constant, which depends on the density of the deployed nodes.

Finally, the rate of change of $P_{ES}$ is quantified using eq. \eqref{eq:rpes}.
\begin{equation}
	\frac{dP_{ES}}{dt}=\theta e^{-\theta(E_{b_{SR}}+E_{b_{RD}})}
\label{eq:rpes}
\end{equation}

\begin{figure}[htb!]
	\centering
	\includegraphics[width=1\linewidth,clip]{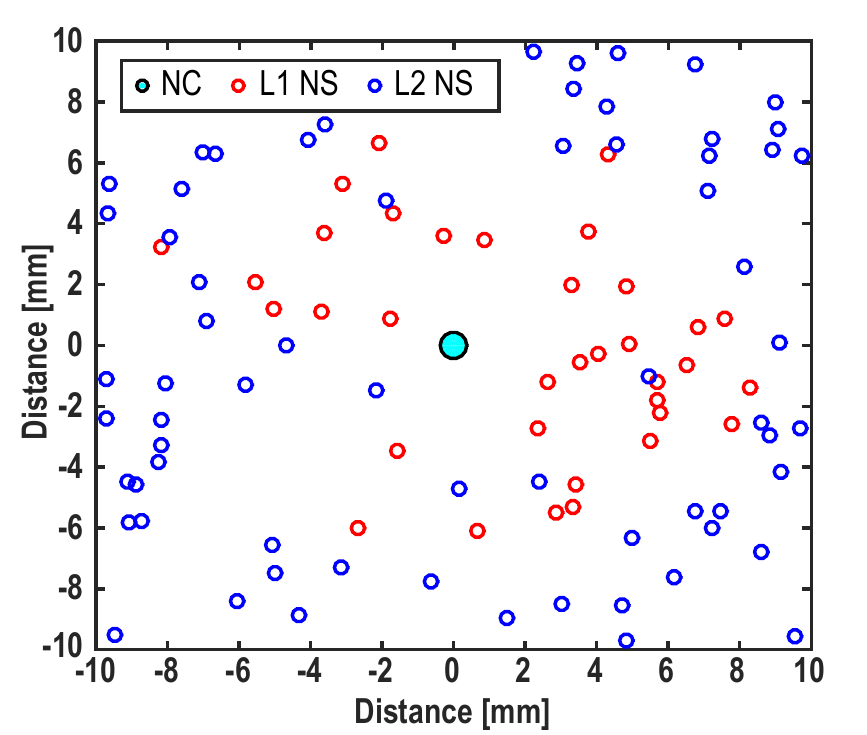}
	\caption{\textbf{Plot of coordinates of deployed nanonodes}. One hundred nanosensors have been deployed using a constant mobility model with a deployment area of $10^{-3} \times 10^{-3} \times 10^{-3} m^3$.}
	\label{ffn}
\end{figure}

\section{Performance Assessment}
\label{sec-perfass}
The model's performance was evaluated for energy efficiency and outage capacity considering the effects of molecular absorption, spreading, path loss, and shadowing. The model's performance was compared to an `Existing' method proposed by Pierobon et al. \cite{17pierobon2014}.

The proposed communication model was simulated using Nano-Sim (see \cite{20piro2013}) for a network with $100$ nanonodes deployed using a constant mobility model within an area of $10$ mm radius with the NC placed at the center. The nanosensor coordinates are shown in Fig.  \ref{ffn}. The nodes send messages through multihop paths to the NC. 
The THz band channel model is devised with a standard gaseous medium comprising $10\%$ of water vapor molecules. Simulation parameters are listed in table \ref{table:1}.

\begin{table}[htb!]
	\renewcommand{\arraystretch}{1.1}
	\caption{Simulation Parameters}
	\label{table:1}
	\begin{center}
        \begin{tabular}{| l  | l  |}
			\hline
			\textbf{Parameter (Unit)} & \textbf{Value}  \\ \hline
			Communication model & EM Thz \\ \hline
			Simulation duration ($s$)& $3$ \\ \hline
			Number of nanonodes & from $500$ to $3000$ \\ \hline
			Deployment area ($m^3$) & $10^{-3} \times 10^{-3} \times 10^{-3} $ \\ \hline
			Node position & Stationary\\ \hline
			Tx Range of nanonodes ($mm$) & $10$ \\ \hline
			Pulse energy ($pJ$) & $100 $ \\ \hline
			Pulse duration ($fs$) & $100 $ \\ \hline
			Pulse interval time ($ps$) & $10 $ \\ \hline
			Packet size ($bits$) & $256$  \\ \hline
			Initial TTL value & $1000$ \\ \hline
			Message generation time interval ($s$) & $0.1 $ \\ \hline
			$C_{nps}$ ($nF$) & $9$\\ \hline
			$V_g$ ($V$) & $0.42 $  \\ \hline
			$\Delta Q$ ($pC$) & $6$  \\ \hline
			Average time between vibrations ($s$) & $\frac{1}{50}$ \\ \hline
			SINR threshold, $\gamma_{th}$ ($dB$)& $12 $   \\ \hline
			Path loss exponent & $3$ \\ \hline
			No. of trials & $10^8$ \\ \hline
			Simulation iteration & 100 \\
			\hline
		\end{tabular}
	\end{center}
\end{table}

\subsection{Energy Usage}
\label{sub-sec-energyusage}
\begin{figure}[htb!]
	\centering
	\includegraphics[width=1\linewidth,clip]{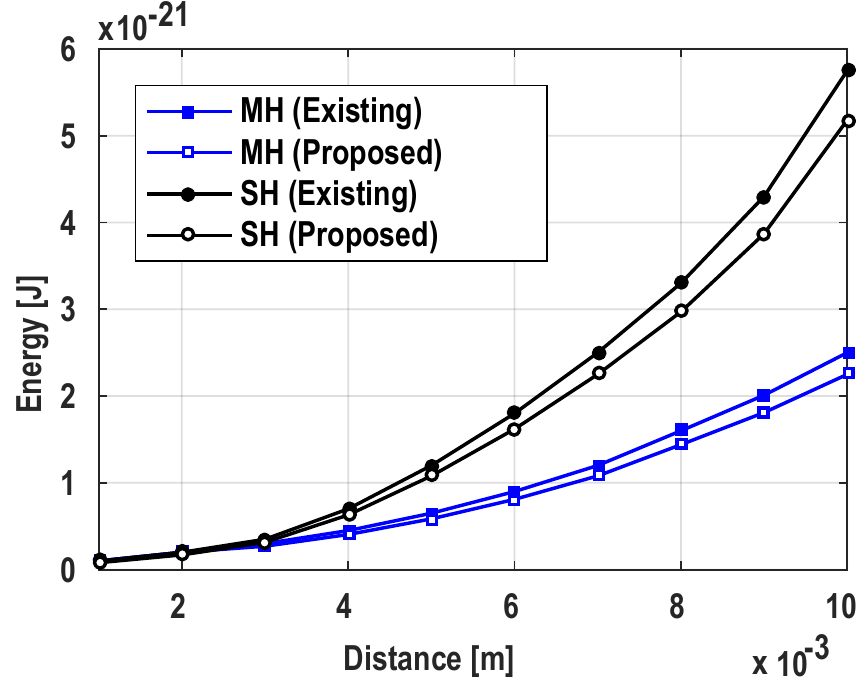}
	\caption{\textbf{Comparison of energy usage}. With increasing distance multihop routing requires less energy than that of singlehop. As the NCC is elected to optimize the energy usage, the proposed routing method operate at a lower energy than that of \cite{17pierobon2014}. }
	\label{ff1}
\end{figure}

Fig.  \ref{ff1} compares the performance of the model in terms of energy spent by each ADTN to transmit packets towards NCC using singlehop and multihop communication. 

As seen in Fig. \ref{ff1}, the energy consumption by an ADTN increases with its distance from the NCC. For a distance less than $\sim3$ mm, the ADTN uses a comparable amount of energy in both singlehop and multihop transmissions. In comparison to a singlehop, multihop transmission of an ADTN requires less energy to transmit to NCC with increasing distances. It also has been found that the proposed algorithm requires less energy, as the NCC is at its best position to collect ADTNs transmission.

\begin{figure}[htb!]
	\centering
	\includegraphics[width=1\linewidth,clip]{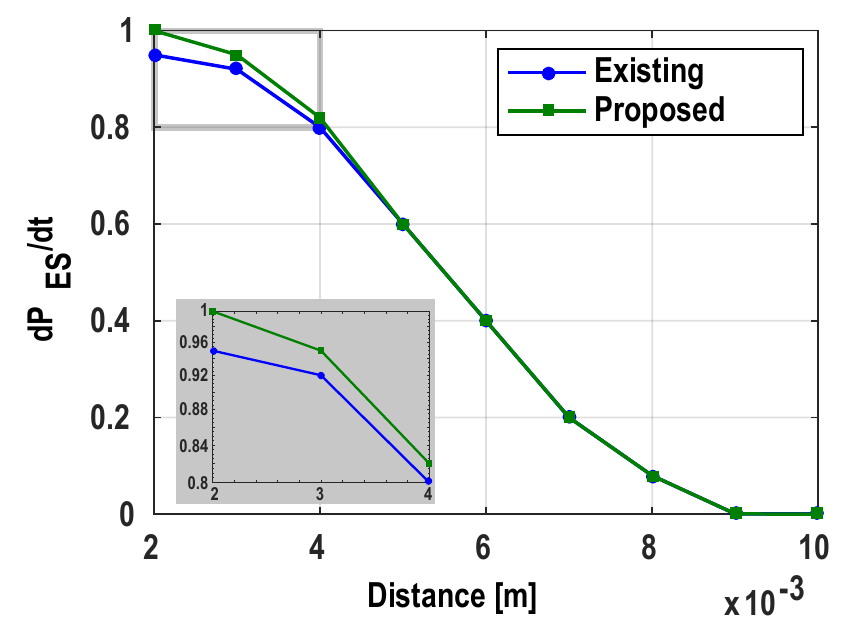}
	\caption{\textbf{Changes in the rate of energy saving probability}. The proposed scheme is able to save more energy for a considerable amount of transmission distance (below $\sim5$ mm) in comparison to the previously proposed one in \cite{17pierobon2014}.}
	\label{ff3}
\end{figure}

As seen in Fig.  \ref{ff3}, in multihop transmission, the rate of change of energy saving probability is higher than that of \cite{17pierobon2014} for average link distances (upto $\sim 5$ mm) between source-relay and relay-destination nodes. This energy usage reduction in the proposed scheme is due to the insightful selection of the NCC based on maximum normalized residual energy among the nodes (see algorithm \ref{algor2-ncca}).

\subsection{Capacity Observation}
\label{sub-sec-capobserv}
\begin{figure}[htb!]
	\centering
	\includegraphics[scale=1]{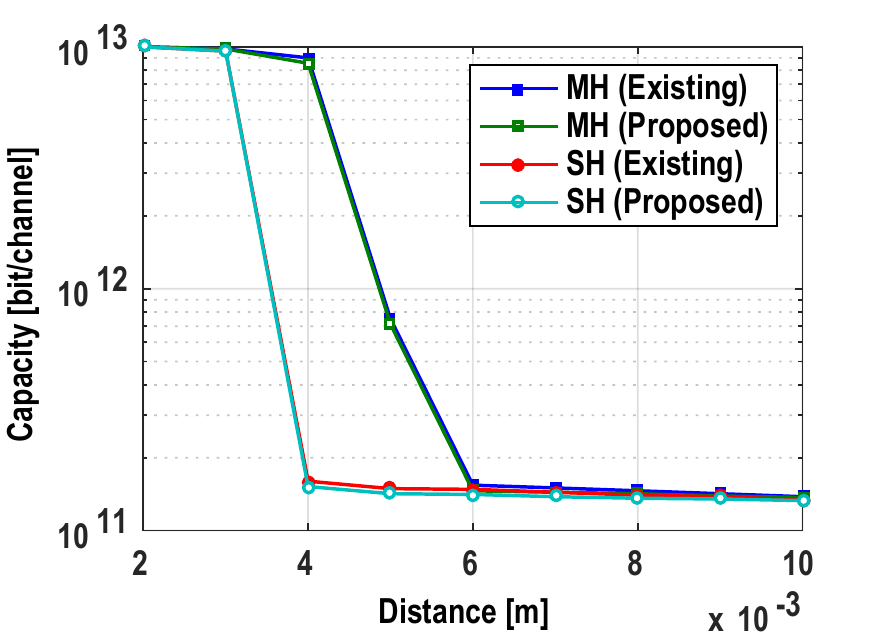}
	
	\caption{\textbf{Channel-wise capacity comparison}. The average capacity decreases with the increase of distance between ADTN and NC.}
	\label{ff2}
\end{figure}

Fig. \ref{ff2} shows the effect of distance between the ADTN and NC on average capacity for singlehop and multihop transmission. The average capacity remains similar for both single and multihop transmissions for upto $\sim3$ mm of distance, after which the capacity of the singlehop decreases in comparison to the multihop until the distance of $\sim6$ mm. For further distances, due to hop-count limits, both transmission modes return to a similar average capacity. It is noteworthy that the proposed routing framework achieves a similar average capacity to the one of Pierobon et al. despite having used less energy (see Fig.  \ref{ff1}). 

\begin{figure}[htb!]
	\centering
	\includegraphics[width=1\linewidth,clip]{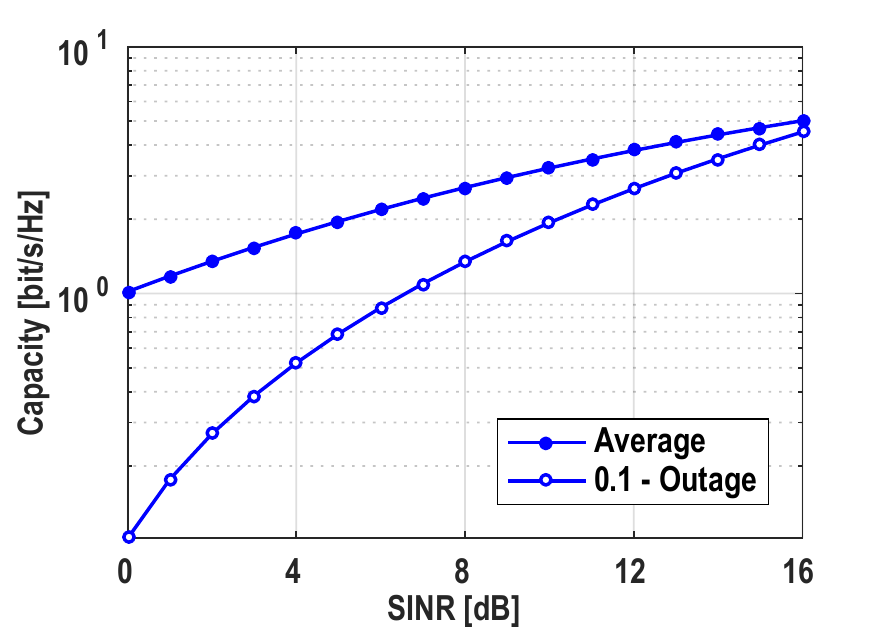}
	\caption{\textbf{Average and outage capacity}. Significant reduction is noticed in the average capacity of the THz communication link considering a 0.1-outage.}
	\label{ffg2}
\end{figure}

It is implicit that the average channel capacity cannot guarantee actual link realization but the outage capacity provides a good estimation (see eq. \eqref{ec1}). For example, the link SINR is low for slow fading and/or shadowing. As expected, with increasing $P_{out}$ the average link capacity decreases (see Fig.  \ref{ffg2}).

\subsection{Outage Probability Analysis}
\label{sub-sec-opa}
The outage probability has been analyzed through Monte-Carlo simulations with parameters $-$ trials: $10^8$, tolerance: $10^{-6}$, $\eta$: $3$, $\mathcal{N}(0,1)$, and \textbf{\textit{k}}: $[1,2,4]$ (see eq. \eqref{eq111} and \eqref{eq7}).

\begin{figure}[htb!]
	\centering
	\includegraphics[width=1\linewidth,clip]{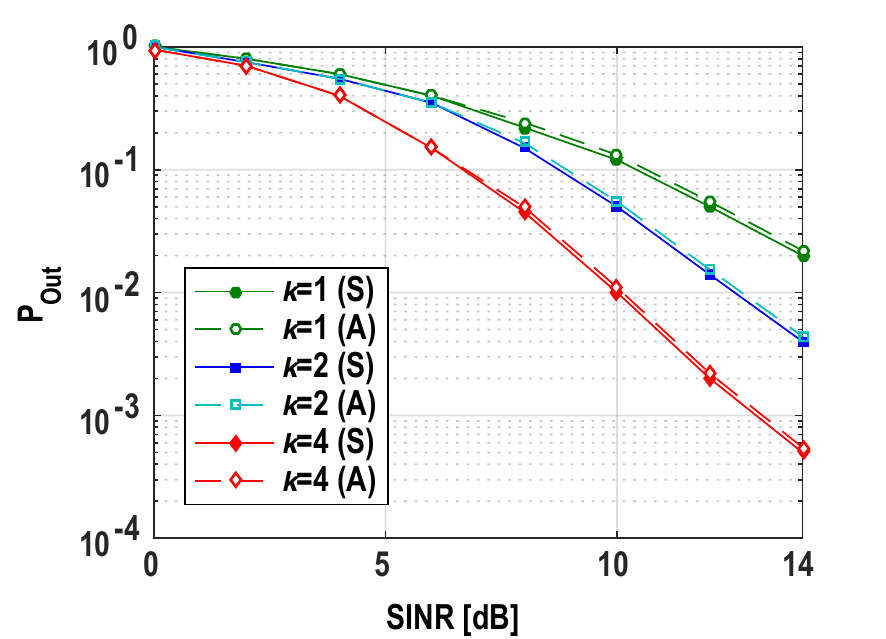}
	\caption{\textbf{Effect of cooperative fusion on $P_{out}$}. In both simulation (S) and analytical (A) results, the outage probability decreases drastically while fusing the information at the fusion node.}
	\label{Fig15}
\end{figure}

The outage probability decreases with an increasing number of links (i.e., \textbf{\textit{k}}). As seen in Fig.  \ref{Fig15}, at $\gamma_j=10$, the outage probabilities are approximately $0.01$ and $0.05$ for $4$ and $2$ parallel links fused at the fusion node (see Fig. \ref{fuse}), and the probability is $0.12$ for a single link. The multisensor fusion at the fusion node enhances the link SINR, which decreases the outage probability and ensures improved quality of service compared to single link multihop communication. The simulation (S) and the analytical (A) results closely match each other ensuring the accuracy of the simulation.

\section{Conclusion}
\label{sec-conc}
Nano electromagnetic communication for WBSN is in its infancy and there are quite a few challenges to be addressed before it can be applied efficiently in the biomedical field. Due to the size of the nanonodes, their resources are limited. Low processing capability, limited battery power, and small memory unit pose major bottlenecks in designing communication protocols for nanonetworks. To overcome the bottlenecks, different protocol stacks have been proposed; however, the network layer has been less studied and explored in nanocommunication.

This work proposed a new energy-efficient forwarding scheme for electromagnetic wireless nanonetworks to improve the performance of nanocommunication over the THz band for a WBSN. The channel behavior of the proposed scheme was investigated considering the combined effect of molecular absorption, spreading loss, and shadowing. To allow continuous operation, the nanonodes of the system require energy harvesting. The proposed scheme facilitates energy efficiency by selecting a multiple layered cluster based multihop communication which is supported by a modified nanosensor energy model. To ensure the quality of service of the WBSN communication, which is hampered by factors such as -- network lifetime, data detection accuracy, and data aggregation delays, we adopted a cooperative fusion approach.

The performance of the scheme was evaluated for energy saving probability, outage probability, and outage capacity. Simulation results demonstrated that the proposed scheme outperforms the existing ones in energy saving and capacity. Additionally, multisensor fusion at the fusion node ensures an enhanced link quality by decreasing the outage probability which is very important in case of smart healthcare applications.

\ifCLASSOPTIONcompsoc
  \section*{Acknowledgments}
\else
  \section*{Acknowledgment}
\fi

This research received no specific grant from any funding agency in the public, commercial, or not-for-profit sectors. The authors are particularly thankful to Mr. Shamim Al Mamun and Ms. Nusrat Jahan for their help and fruitful discussion.

\ifCLASSOPTIONcaptionsoff
  \newpage
\fi

\bibliographystyle{IEEEtran}
\bibliography{ref}

\begin{IEEEbiographynophoto}{Fariha Afsana} 
(GSM'11) received her B.Sc (Honors.) and M.Sc degree in Information Technology from Jahangirnagar University, Savar, Dhaka, Bangladesh in 2015 and 2016 respectively. Mrs. Afsana’s research interests include nanotechnology, next generation wireless networks, cognitive radio networks, cooperative networks and wireless sensor networks. She is a Graduate Student member of IEEE.
\end{IEEEbiographynophoto}

\vspace{-2.5em}

\begin{IEEEbiographynophoto}{Md. Asif Ur Rahman} 
received his B.Sc. and M.Sc. degree from American International University Bangladesh (AIUB). During his junior year he had participated and ranked in many national and international programming contests. After graduating, he joined the Department of Computer Science at AIUB as a faculty member. Currently, he is serving as Assistant Professor. He also retains real-world experience in software design and development from leading industries of Bangladesh. His research interests include computer vision, artificial intelligence, network security and simulation, IoT and IoNT.
\end{IEEEbiographynophoto}

\vspace{-2.5em}

\begin{IEEEbiographynophoto}{Muhammad R. Ahmed} 
currently serves as Seiner Lecturer at Radar and Radio Communications, Marine Engineering Department, Military Technology College, Muscat, Oman- (University of Portsmouth, United Kingdom). He worked as Lecturer at the Faculty of Information Sciences and Engineering, University of Canberra (UC) and as a research officer at Australian National University (ANU), Australia. He was a distinguished member of the Board of directors of ITE\&E, Engineers Australia in 2011. He obtained his PhD at the UC, Australia. He has received a M.Eng. in Telecommunication and a M.Eng. Management degree from the University of Technology, Sydney (UTS), Australia. He obtained his B.Eng. (Hons) Electronics Major in Telecommunications degree from Multimedia University (MMU), Malaysia. Dr. Muhammad has also authored several papers in Wireless Sensor Networks, Distributed Wireless Communication, Blind Source Separation, RF technologies, and RFID implementation. He has published many papers in high level of journals and conferences.
\end{IEEEbiographynophoto}

\vspace{-2.5em}

\begin{IEEEbiographynophoto}{Mufti Mahmud}
(GSM'08, M'11, SM'16) originally from Bangladesh, received his post-school education in India (B.Sc. from University of Madras and M.Sc. from University of Mysore, both in Computer Science, in 2001 and 2003 respectively) and Italy (M.S. in Bionanotechnology from University of Trento and Ph.D. in Information Engineering from University of Padova in 2008 and 2011 respectively). A recipient of Marie-Curie Fellowship, Dr. Mahmud served at various positions in the industry and academia in India, Bangladesh, Italy, and Belgium. With over 60 publications in leading Journals and Conferences, Dr. Mahmud is a Neuroinformatics expert. He has released two open source toolboxes (SigMate, and QSpike Tools) for processing and analysis of brain signals. Additionally, Dr. Mahmud's expertise includes Healthcare and Neuroscience Big Data analytics, Advanced Machine Learning applied to Biological data, Assistive Brain-Machine Interfacing, Computational Neuroscience, Personalized and Preventive [e/m]-Healthcare, Internet of Healthcare Things, Cloud Computing, Security and Trust Management in Cyber-Physical Systems, and Crowd Analysis. Dr. Mahmud serves as Associate Editor of Springer-Nature's Cognitive Computation journal, successfully organized numerous special sessions in leading conferences, served many internationally reputed conferences in different capacities (e.g., programme, Organization \& advisory Committee member), and also served as referee for many high-impact journals.
\end{IEEEbiographynophoto}

\vspace{-2.5em}

\begin{IEEEbiographynophoto}{M Shamim Kaiser}
(SM'16) received  the B.Sc. (Honors) and M.S. degrees in Applied Physics Electronics and Communication Engineering from the University of Dhaka, Bangladesh 2002 and 2004 respectively, and a Ph.D. in Telecommunications from the Asian Institute of Technology (AIT) Pathumthani, Thailand, in 2010. He is working as Associate Professor in the Institute of Information Technology of Jahangirnagar University, Dhaka, Bangladesh. His current research interests include Multi-hop Wireless Networks, Big Data Analytics and Cyber Security. Dr. Kaiser is a Life Member of the Bangladesh Electronic Society, Bangladesh Physical Society, and Bangladesh Computer Society. He is a senior member of IEEE, USA, a member of IEICE, Japan and ExCom member of IEEE Bangladesh Section.
\end{IEEEbiographynophoto}

\vfill

\end{document}